\documentclass[preprint,12pt,3p]{elsarticle}
\usepackage{graphicx}
\usepackage{dcolumn}
\usepackage{amssymb}
\usepackage{epsfig}
\usepackage{amsthm}
\usepackage{amsmath}
\usepackage{mathrsfs}
\usepackage{bm}

\newcommand{\ket}[1]{\mathrm{\left| #1 \right>}\,}
\newcommand{\abs}[1]{\left| #1 \right|}
\begin{document}   
\journal{Chemical Physics}

\title{Exact dynamics of a Gaussian wave-packet in two potential curves coupled at a point}

\author[label1]{Saravanan Rajendran\corref{cor1}}

\cortext[cor1]{Corresponding author}
\ead{saravanan\_p@students.iitmandi.ac.in}
\ead[url]{http://saravananrajendran.weebly.com}

\author[label2]{Aniruddha Chakraborty}
\fntext[label1,label2]{School of Basic Sciences, Indian Institute of Technology Mandi, Kamand, Himachal Pradesh-175005, India}
\ead{aniruddha\_oregon@yahoo.com}

\begin{frontmatter}
\begin{abstract}
We present a method to calculate exact dynamics of a wave-packet in a quantum two-state problem with Dirac delta coupling. The advantage of our method is that the calculations are done in the time domain. Hence inverting the solutions from other domains (Laplace or Fourier domain) do not intervene us from presenting the solution in the time domain. The initial wave packet is considered to be a Gaussian function. The wave propagation in the quantum state is governed by time-dependent Schr\"odinger equation and the coupling is given by non-diagonal element in the matrix representation. We present the exact analytic forms of the wave function for both the states in the time domain.
\end{abstract}

\date{\today}
\end{frontmatter}

\section{Introduction}
Non-adiabatic transitions in intra- and intermolecular processes were understood by the multistate models following from the work of Landau\cite{landau1932theory}. The notion is an interdisciplinary concept and has applications in understanding many processes is science, namely, electronic transitions\cite{nikitin1999nonadiabatic,child1996molecular,thiel1990landau,nakamura1991basic}, neutrino oscillations\cite{kim1987adiabatic}, buckling state transition in nanorods as a two-state process\cite{chakraborty2011buckled}, in Stern-Gerlach experiment\cite{daghigh2016consequence}, dissociation reaction as a curve crossing effect\cite{chakraborty2011predissociation,chakraborty2011curve} and in many other fields\cite{nakamura2002nonadiabatic,nikitin1999nonadiabatic,devault1984quantum,shaik1991zb}. Many scattering processes are also theoretically similar to that of spectroscopic processes where the photodissociation cross section is expressed in terms of the scattering matrix\cite{nakamura1991basic}. Since then, there were many attempts to solve the transition probability between the states\cite{zener1932non,stuckelberg1932theory,landau1932theorie}. There were some recent works involved in solving the curve crossing models where the coupling was a single Dirac delta coupling\cite{chakraborty2013multi,chakraborty2009effect}, arbitrary coupling as a collection of Dirac delta coupling \cite{chakraborty2015transfer,diwaker2012curve}, time varied strength\cite{panda2016exact} and position\cite{chakraborty2013non} Dirac delta couplings.The multistate problems are being solved only for the case of Dirac delta function analytically while solving other couplings still remain an open problem. The above works involve the calculation in Laplace domain and require the pre-knowledge of the exact functional form of the Green's function. Hence, the exact time-dependent wavefunctions are not expressed in many cases. In this paper, we consider two quantum states whose wavepacket propagation is given by Schr\"odinger equation with the presence of coupling as written as an equation,
\begin{equation}
\left( \begin{array}{cc}
\hat{H_{11}} & V_{12} \\
V_{21} & \hat{H_{22}}\\
\end{array}\right)
\left( \begin{array}{c}
\Psi_1(x,t) \\
\Psi_2(x,t)\\
\end{array}\right)
=i\hbar\left( \begin{array}{c}
\frac{\partial \Psi_1}{\partial t} \\
\frac{\partial \Psi_2}{\partial t} \\
\end{array}\right)
\end{equation}
The coupling between the states is given by a Dirac delta function of arbitrary strength $[V_{12}(x)=V_{21}(x)=k_0\delta(x-x_c)]$ and the states are given by two flat potential curves $[V_2(x)-V_1(x)=V_0]$. Owing to the translational symmetry in the potential, we put the coupling in the origin to reduce the number of parameters involved. Initially, the wavepacket is assumed to be in the ground state which is peaked at initial position $x_0$. The wavepacket is assumed to take the Gaussian functional form,
$$\Psi_1(x,0)=\left(\frac{1}{2\pi\sigma^2}\right)^{1/4}e^{-\frac{(x+x_0)^2}{4\sigma^2}+ik_1(x+x_0)}$$
with $\sigma$ is the measure of width and $k_1$ is the momentum of the wavepacket in the given state. Our aim is to calculate for the time-dependent wavefunction in both these states separately. We organise our paper in sections in which section II gives the methodology and calculation involved in solving the concerned problem. In section III, we present the exact analytical results, asymptotic solutions and discussions. In section IV, we conclude our paper stating the possible considerations and the scope of the present work.
\section{Methodology : Kernel method to calculate wavepacket dynamics}
The two quantum states coupled through a finite strength Dirac delta function is written as,
\begin{equation}-\frac{\hbar^2}{2m}\frac{\partial^2}{\partial x^2}\Psi_1+k_0\delta(x)\Psi_2=i\hbar\frac{\partial \Psi_1}{\partial t}
\label{eqn:ground}
\end{equation}
\begin{equation}-\frac{\hbar^2}{2m}\frac{\partial^2}{\partial x^2}\Psi_2+V_0\Psi_2+k_0\delta(x)\Psi_1=i\hbar\frac{\partial \Psi_2}{\partial t}
\label{eqn:excited}
\end{equation}
Seeking for the stationary state solutions for the Eqs. \eqref{eqn:ground}\&\eqref{eqn:excited} under separable approximation, $\lim_{t\to\infty}\Psi_i(x,t)=\phi_i(x)T(t)$ yields,
\begin{equation}-\frac{\hbar^2}{2m}\frac{\partial^2}{\partial x^2}\phi_1(x)+k_0\delta(x)\phi_2(x)=E\phi_1(x)
\label{station1}
\end{equation}
\begin{equation}-\frac{\hbar^2}{2m}\frac{\partial^2}{\partial x^2}\phi_2(x)+V_0 \phi_2(x)+k_0\delta(x)\phi_1(x)=E\phi_2(x)
\label{station2}
\end{equation}
$$i\hbar\frac{\partial T}{\partial t}=ET\implies\hat{T}=e^{-iH_it/\hbar}$$
The general solutions of Eqs. \eqref{eqn:ground} \& \eqref{eqn:excited} under physical arguments is given by,
$$\phi_1(x)=
\begin{cases}
Ae^{ikx}+Be^{-ikx},~ x<0\\
 Ce^{ikx},~~~~~~~  ~~~~     x>0,~~~~~~~ \text{where}~~k =\sqrt{\frac{2mE}{\hbar^2}}
\end{cases}$$
$$\phi_2(x)=
\begin{cases}
De^{-ik'x},& x<0\\
Ee^{ik'x}, &    x>0,~~~~~~~ \text{where}~~k'=\sqrt{\frac{2m(E-V_0)}{\hbar^2}}
\end{cases}$$
Now the effect of Dirac delta function is incorporated in course of boundary conditions at the boundary $x=0$,

$$(\phi_{1})_{0+\epsilon}=(\phi_{1})_{0-\epsilon}$$
$$\left(\frac{\partial \phi_1}{\partial x}\right)^{0+\epsilon}_{0-\epsilon}=2mk_0\phi_2(0)/\hbar^2$$
$$(\phi_{2})_{0+\epsilon}=(\phi_{2})_{0-\epsilon}$$
$$\left(\frac{\partial \phi_2}{\partial x}\right)^{0+\epsilon}_{0-\epsilon}=2mk_0\phi_1(0)/\hbar^2$$
$$\phi_1(x)=
\begin{cases}
e^{ikx}+\left(\frac{m^2k_0^2}{-kk'\hbar^4-m^2k_0^2}\right)e^{-ikx},& x<0\\
 e^{ikx}+
\left(\frac{m^2k_0^2}{-kk'\hbar^4-m^2k_0^2}\right)e^{ikx},&   x>0,
\end{cases}$$

$$\phi_2(x)=
\begin{cases}
\frac{\hbar^2(ik)m^2k_0^2}{mk_0(-kk'\hbar^4-m^2k_0^2)}e^{-ik'x},& x<0\\
\frac{\hbar^2(ik)m^2k_0^2}{mk_0(-kk'\hbar^4-m^2k_0^2)}e^{ik'x}, &    x>0,~~~~~~~ \text{where}~~k'=\sqrt{\frac{2m(E-V_0)}{\hbar^2}}
\end{cases}$$
Taking the Fourier transform of the initial wave function,
$$\tilde \Psi(k,0)=\frac{1}{\sqrt{2\pi}}\int_{-\infty}^{\infty} \! \Psi_1(x,0)e^{-ikx} \, \mathrm{d}x$$
$$\tilde \Psi(k,0)=\left(\frac{2\sigma^2}{\pi}\right)^{1/4}e^{-\sigma^2(k-k_1)^2+ikx_0}$$
The initial wavefunction in the Fourier representation is written as,
$$\Psi_1(x,0)=\left(\frac{\sigma^2}{2\pi^3}\right)^{1/4}\int_{-\infty}^{\infty} \!e^{-\sigma^2(k-k_1)^2+ikx_0} e^{ikx} \, \mathrm{d}k$$
We know the time operator obtained from Taylor's series expansion,
$$\hat{T}=e^{-iH_it/\hbar}$$
$$\Psi(x,t)=\hat{T}\Psi_1(x,0)$$
The time evolution of the wave packet corresponding to unscattered wave part of the stationary solution,
$$\Psi_1^{(1)}(x,t)=\left(\frac{\sigma^2}{2\pi^3}\right)^{1/4}\int_{-\infty}^{\infty} \!e^{-iH_1t/\hbar}e^{-\sigma^2(k-k_1)^2+ikx_0} e^{ikx} \, \mathrm{d}k$$
$e^{ikx}$ is an eigenfunction of the operator $\hat{H_1}$, the time evolution turns out to be,
$$\Psi_1^{(1)}(x,t)=\left(\frac{\sigma^2}{2\pi^3}\right)^{1/4}\int_{-\infty}^{\infty} \!e^{-ihk^2t/{2m}}e^{-\sigma^2(k-k_1)^2+ikx_0} e^{ikx} \, \mathrm{d}k$$


$$\Psi_1^{(1)}(x,t)=\frac{1}{(2\pi)^{1/4}}\sqrt{\frac{\sigma}{(\sigma^2+\frac{i\hbar t}{2m})}}e^{\frac{-(x+x_0)^2}{4(\sigma^2+\frac{i\hbar t}{2m})}-\frac{i\sigma^2}{(\sigma^2+\frac{i\hbar t}{2m})}(k_1(x+x_0)-\frac{\hbar k_1^2 t}{2m})}$$
Calculation of the scattered wavepacket is given by the initial wave propagated with scattered part kernel of the stationary solution,
$$\Psi_1^{(2)}(x,0)=\left(\frac{\sigma^2}{2\pi^3}\right)^{1/4}\int_{-\infty}^{\infty} \!e^{-\sigma^2(k-k_1)^2+ikx_0}\left(\frac{m^2k_0^2}{-kk'\hbar^4-m^2k_0^2}\right)e^{ik\abs x} \, \mathrm{d}k$$
$$\Psi_1^{(2)}(x,t)=\left(\frac{\sigma^2}{2\pi^3}\right)^{1/4}\int_{-\infty}^{\infty} \!e^{-\sigma^2(k-k_1)^2+ikx_0}\left(\frac{m^2k_0^2}{-kk'\hbar^4-m^2k_0^2}\right)e^{-iH_1t/\hbar}e^{ik\abs x} \, \mathrm{d}k$$
Completing squares in the integrand,
\begin{equation}
\Psi_1^{(2)}(x,t)=\left(\frac{\sigma^2}{2\pi^3}\right)^{1/4}(-\frac{m^2k_0^2}{\hbar^4})e^{-\sigma^2k_1^2}e^{-\frac{\left(\abs{x}+x_0-i\sigma^22k_1\right)^2}{4(\sigma^2+i\hbar t/2m)}}\int_{-\infty}^{\infty} \!\frac{e^{-(\sigma^2+i\hbar t/2m)\left[k-\frac{i\left(\abs{x}+x_0-i\sigma^22k_1\right)}{2(\sigma^2+i\hbar t/2m)}\right]^2}}{kk'+\frac{m^2k_0^2}{\hbar^4}} \, \mathrm{d}k
\label{psi2}
\end{equation}
$$I=\int_{-\infty}^{\infty} \!\frac{e^{-(\sigma^2+i\hbar t/2m)\left[k-\frac{i\left(\abs{x}+x_0-i\sigma^22k_1\right)}{2(\sigma^2+i\hbar t/2m)}\right]^2}}{k\sqrt{k^2-\frac{2mV_0}{\hbar^2}}+\frac{m^2k_0^2}{\hbar^4}} \, \mathrm{d}k$$

$$I=\int_{-\infty}^{\infty} \!\frac{\left(k\sqrt{k^2-\frac{2mV_0}{\hbar^2}}-\frac{m^2k_0^2}{\hbar^4}\right)e^{-(\sigma^2+i\hbar t/2m)\left[k-\kappa(x,t)\right]^2}}{\left(k\sqrt{k^2-\frac{2mV_0}{\hbar^2}}\right)^2-\left(\frac{m^2k_0^2}{\hbar^4}\right)^2} \, \mathrm{d}k$$
with $\kappa(x,t)=\frac{i\left(\abs{x}+x_0-i\sigma^22k_1\right)}{2(\sigma^2+i\hbar t/2m)}$. The poles of k are,
$$k_{\pm\pm}=\pm\sqrt{\frac{mV_0}{\hbar^2}\pm\sqrt{\left(\frac{mV_0}{\hbar^2}\right)^2+\left(\frac{m^2k_0^2}{\hbar^4}\right)^2}}$$
$$I=\int_{-\infty}^{\infty} \!\frac{\left(k\sqrt{k^2-\frac{2mV_0}{\hbar^2}}-\frac{m^2k_0^2}{\hbar^4}\right)e^{-(\sigma^2+i\hbar t/2m)\left[k-\kappa(x,t)\right]^2}}{\prod_{i=1}^{4}(k-k_i)} \, \mathrm{d}k$$
The indices 1,2,3,4 are $-+$, $++$, $--$, $+-$ respectively. The poles 1 and 2 necessarily lie on the real line whereas the poles 3 and 4 lies in the complex plane depending on the real values of $V_0$ $k_0$
\begin{figure}[h]
    \centering
    \includegraphics[scale=0.8]{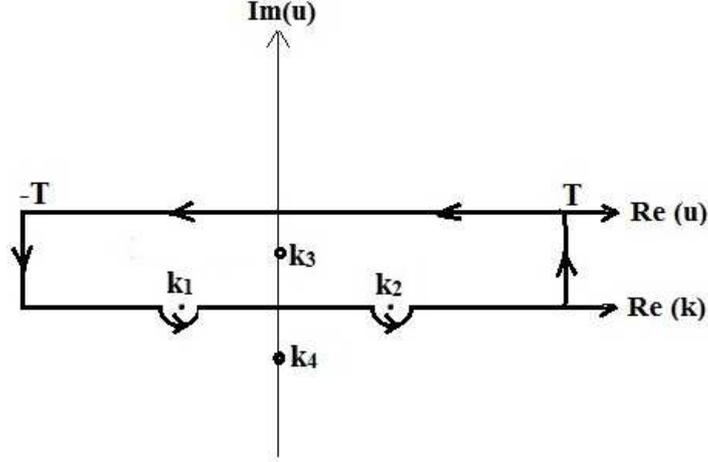}
   \caption{Schematic representation of the Integrand in the complex plane made by k-space. The poles $k_3$ and $k_4$ are dynamic, that they remain equal and opposite in the plane with respect to origin depending on the time.}
    \label{quantum contour}
\end{figure}
\begin{eqnarray}
\oint \!I=\left(\int_{-T}^{-\epsilon_1-k_1}+\int_{\epsilon_1-k_1}^{-\epsilon_2-k_2}+\int_{\epsilon_2-k_2}^{T}\right)\!\frac{\left(k\sqrt{k^2-\frac{2mV_0}{\hbar^2}}-\frac{m^2k_0^2}{\hbar^4}\right)e^{-(\sigma^2+i\hbar t/2m)(k-\kappa(x,t))^2}}{\prod_{i=1}^{4}(k-k_i)} \, \mathrm{d}k \\ \nonumber
+\sum_{j=1}^{2}\lim_{\epsilon_j\to0}\int_{\pi}^{0}\frac{\left((\epsilon_je^{i\theta}-k_j)\sqrt{(\epsilon_je^{i\theta}-k_j)^2-\frac{2mV_0}{\hbar^2}}-\frac{m^2k_0^2}{\hbar^4}\right)e^{-(\sigma^2+i\hbar t/2m)((\epsilon_je^{i\theta}-k_j)-\kappa(x,t))^2}}{\prod_{i=1}^{4}((\epsilon_je^{i\theta}-k_j)-k_i)}\epsilon_j(ie^{i\theta}) \, \mathrm{d}\theta\\ +\int_{T}^{-T}\frac{\left((u+\kappa)\sqrt{(u+\kappa)^2-\frac{2mV_0}{\hbar^2}}-\frac{m^2k_0^2}{\hbar^4}\right)e^{-(\sigma^2+i\hbar t/2m)u^2}}{\prod_{i=1}^{4}((u+\kappa)-k_i)} \, \mathrm{d}u \nonumber \\+\left(\int_{T}^{T+\kappa}+\int_{-T+\kappa}^{-T}\right)\frac{\left(k+iIm(k)\sqrt{k+iIm(k)^2-\frac{2mV_0}{\hbar^2}}-\frac{m^2k_0^2}{\hbar^4}\right)e^{-(\sigma^2+i\hbar t/2m)(k+iIm(k)-\kappa(x,t))^2}}{\prod_{i=1}^{4}(k+iIm(k)-k_i)} \, \mathrm{d}Im(k)\nonumber
\end{eqnarray}
the term II goes to zero as the contour is around a point of radius tending to zero and the term IV goes to zero because the integrand cancels with the component that is equal and opposite sign. In the limits $T\to\infty$ and $\epsilon_j\to0$,
\begin{eqnarray}
\oint \!I=\int_{-\infty}^{\infty}\!\frac{\left(k\sqrt{k^2-\frac{2mV_0}{\hbar^2}}-\frac{m^2k_0^2}{\hbar^4}\right)e^{-(\sigma^2+i\hbar t/2m)(k-\kappa(x,t))^2}}{\prod_{i=1}^{4}(k-k_i)} \, \mathrm{d}k 
\\ +\int_{\infty}^{-\infty}\frac{\left((u+\kappa)\sqrt{(u+\kappa)^2-\frac{2mV_0}{\hbar^2}}-\frac{m^2k_0^2}{\hbar^4}\right)e^{-(\sigma^2+i\hbar t/2m)u^2}}{\prod_{i=1}^{4}((u+\kappa)-k_i)} \, \mathrm{d}u \nonumber
\end{eqnarray}
\begin{equation}
    \oint \!I=2\pi i\sum_{l=1}^4\mu(k_l)\frac{\left(k_l\sqrt{k_l^2-\frac{2mV_0}{\hbar^2}}-\frac{m^2k_0^2}{\hbar^4}\right)e^{-(\sigma^2+i\hbar t/2m)(k_l-\kappa(x,t))^2}}{\prod_{i\neq l,i=1}^{4}(k_l-k_i)}
\end{equation}
$$\mu(k_l)=\begin{cases}
-1  ~~~~~~~~~~~k_l\in [-\frac{f(x)^2}{2(\sigma^2+\frac{i\hbar t}{2m})},0].\\
1 ~~~~~~~~~~~k_l\in [0,\frac{f(x)^2}{2(\sigma^2+\frac{i\hbar t}{2m})}].\\
0 ~~~~~~~~~~~~k_l\notin [-\frac{f(x)^2}{2(\sigma^2+\frac{i\hbar t}{2m})},\frac{f(x)^2}{2(\sigma^2+\frac{i\hbar t}{2m})}].
\end{cases}$$
\begin{eqnarray}
I=2\pi i\sum_{l=1}^4\mu(k_l)\frac{\left(k_l\sqrt{k_l^2-\frac{2mV_0}{\hbar^2}}-\frac{m^2k_0^2}{\hbar^4}\right)e^{-(\sigma^2+i\hbar t/2m)(k_l-\kappa(x,t))^2}}{\prod_{i\neq l,i=1}^{4}(k_l-k_i)}\\ +\int_{-\infty}^{\infty}\frac{\left((u+\kappa)\sqrt{(u+\kappa)^2-\frac{2mV_0}{\hbar^2}}-\frac{m^2k_0^2}{\hbar^4}\right)e^{-(\sigma^2+i\hbar t/2m)u^2}}{\prod_{i=1}^{4}((u+\kappa)-k_i)} \, \mathrm{d}u\nonumber
\end{eqnarray}

\begin{eqnarray}
\Psi_1^{(2)}(x,t)=B(x,t)[2\pi i\sum_{l=1}^4\mu(k_l)\frac{\left(k_l\sqrt{k_l^2-\frac{2mV_0}{\hbar^2}}-\frac{m^2k_0^2}{\hbar^4}\right)e^{-(\sigma^2+i\hbar t/2m)(k_l-\kappa(x,t))^2}}{\prod_{i\neq l,i=1}^{4}(k_l-k_i)}\\ \nonumber+\int_{-\infty}^{\infty}\frac{\left((u+\kappa)\sqrt{(u+\kappa)^2-\frac{2mV_0}{\hbar^2}}-\frac{m^2k_0^2}{\hbar^4}\right)e^{-(\sigma^2+i\hbar t/2m)u^2}}{\prod_{i=1}^{4}((u+\kappa)-k_i)} \, \mathrm{d}u]
\label{psi2}
\end{eqnarray}
with $B(x,t)=\left(\frac{\sigma^2}{2\pi^3}\right)^{1/4}(-\frac{m^2k_0^2}{\hbar^4})e^{-\sigma^2k_1^2}e^{-\frac{\left(\abs{x}+x_0-i\sigma^22k_1\right)^2}{4(\sigma^2+i\hbar t/2m)}}$.
The wave propagation in the other state can be obtained by solving the integral that is given,
$$\Psi_2(x,t)=\left(\frac{\sigma^2}{2\pi^3}\right)^{1/4}\int_{-\infty}^{\infty} \!e^{-\sigma^2(k-k_1)^2+ikx_0}\frac{\hbar^2(ik)m^2k_0^2}{mk_0(-kk'\hbar^4-m^2k_0^2)}e^{-iH_2t/\hbar}e^{ik'\abs{x}} \, \mathrm{d}k$$

\section{Asymptotic solutions and discussions}
The total wavepacket in the ground state is given by,
$$\Psi_1(x,t)=\Psi_1^{1}(x,t)+\Psi_1^{2}(x,t)$$
\begin{eqnarray}
\Psi_1(x,t)=Ae^{\frac{-(x+x_0)^2}{4(\sigma^2+\frac{i\hbar t}{2m})}-\frac{i\sigma^2}{(\sigma^2+\frac{i\hbar t}{2m})}(k_1(x+x_0)-\frac{\hbar k_1^2 t}{2m})}+\\ B(x,t)[2\pi i\sum_{l=1}^4\mu(k_l)\frac{\left(k_l\sqrt{k_l^2-\frac{2mV_0}{\hbar^2}}-\frac{m^2k_0^2}{\hbar^4}\right)e^{-(\sigma^2+i\hbar t/2m)(k_l-\kappa(x,t))^2}}{\prod_{i\neq l,i=1}^{4}(k_l-k_i)}\\\nonumber \nonumber+\int_{-\infty}^{\infty}\frac{\left((u+\kappa)\sqrt{(u+\kappa)^2-\frac{2mV_0}{\hbar^2}}-\frac{m^2k_0^2}{\hbar^4}\right)e^{-(\sigma^2+i\hbar t/2m)u^2}}{\prod_{i=1}^{4}((u+\kappa)-k_i)} \, \mathrm{d}u]
\label{psi2}
\end{eqnarray}
with $A=\frac{1}{(2\pi)^{1/4}}\sqrt{\frac{\sigma}{(\sigma^2+\frac{i\hbar t}{2m})}}$.
And the solution of the second state $\ket{2}$ is given by,
$$\Psi_2(x,t)=\left(\frac{\sigma^2}{2\pi^3}\right)^{1/4}\int_{-\infty}^{\infty} \!e^{-\sigma^2(k-k_1)^2+ikx_0}\frac{\hbar^2(ik)m^2k_0^2}{mk_0(-kk'\hbar^4-m^2k_0^2)}e^{-iH_2t/\hbar}e^{ik'\abs{x}} \, \mathrm{d}k$$
As th integrals are not exactly solvable, we take two asymptotic cases : The cases are 
\begin{itemize}
    \item When the energy of the wavepacket is much greater than the separation between the potential curves $E>>V_0$
    \item When the energy of the wavepacket is much greater than the separation between the potential curves $E<<V_0$

\end{itemize}
For case 1, where $E>>V_0$, the $k'\approx k$, the integral \eqref{psi2} turns as Goodwin-Staton integral with two simple roots. Solutions for arbitrary $k_0$ and higher $k_0$ are also present. We present her the solution for $k_0\to\infty$ for simplicity. For detailed solution, please refer to our earlier publications\cite{stat2state}.
With mappings:\\
$-\frac{-\epsilon_0^2}{2\pi}e^{\frac{-(\abs{x}+x_0)}{4(\sigma^2+Dt)}}=B$, $D=\frac{i\hbar}{2m}$,$\frac{i(\abs{x}+x_0)}{2(\sigma^2+Dt)}=\frac{i(\abs{x}+x_0-i\sigma^2k_1)}{2(\sigma^2+\frac{i\hbar}{2m}t)}$,$\epsilon_0=\frac{m^2k_0^2}{\hbar^4}$;
\begin{eqnarray}
\Psi_1(x,t)=Ae^{\frac{-(x+x_0)^2}{4(\sigma^2+\frac{i\hbar t}{2m})}-\frac{i\sigma^2}{(\sigma^2+\frac{i\hbar t}{2m})}(k_1(x+x_0)-\frac{\hbar k_1^2 t}{2m})}\\ \nonumber+B\sum_{n=0}^\infty\frac{(-1)^n(2n-1)!!(\hbar^4)^{2n+2}}{(m^2k_0^2)^{2n+2)}2^n(\sigma^2+\frac{i\hbar t}{2m})^{n+1/2}}\sqrt{\pi}
\label{psi1_E>V}
\end{eqnarray}
$$\Psi_2(x,t)=B\frac{i(\abs{x}+x_0-i\sigma^2k_1)}{2(\sigma^2+\frac{i\hbar}{2m}t)}\sum_{n=0}^\infty\frac{(-1)^n(2n-1)!!(\hbar^4)^{2n+2}}{(m^2k_0^2)^{2n+2)}2^n(\sigma^2+\frac{i\hbar t}{2m})^{n+1/2}}\sqrt{\pi}$$
For case 2, k' is effectively decomposed as $k'=k-\sqrt{\frac{2mV_0}{\hbar^2}}$.The solutions are presented in our earlier publications\cite{stat1state},
With mappings:\\

$-\frac{-i\epsilon_0}{2\pi}e^{\frac{-(\abs{x}+x_0)}{4(\sigma^2+Dt)}}=\frac{B}{i\sqrt{\frac{2mV_0}{\hbar^2}}}$, $D=\frac{i\hbar}{2m}$,$\frac{i(\abs{x}+x_0)}{2(\sigma^2+Dt)}=\frac{i(\abs{x}+x_0-i\sigma^2k_1)}{2(\sigma^2+\frac{i\hbar}{2m}t)}$
$\epsilon_0=-\frac{m^2k_0^2}{\hbar^4}\sqrt{\frac{2mV_0}{\hbar^2}}$;
\begin{eqnarray}
\Psi_1(x,t)=Ae^{\frac{-(x+x_0)^2}{4(\sigma^2+\frac{i\hbar t}{2m})}-\frac{i\sigma^2}{(\sigma^2+\frac{i\hbar t}{2m})}(k_1(x+x_0)-\frac{\hbar k_1^2 t}{2m})}\\ \nonumber+B\sum_{n=2s,s=0}^\infty\frac{(n-1)!!(\hbar^5)^{n+1)}}{(-im^2k_0^2\sqrt{2mV_0})^{2(n+2)}2^{n/2}(\sigma^2+\frac{i\hbar t}{2m})^{n+1/2}}\sqrt{\pi}
\label{psi1_E>V}
\end{eqnarray}
\begin{eqnarray}
\Psi_2(x,t)=Be^{\sqrt{\frac{-2mV_0\abs{x}}{\hbar^2}}}[\frac{i(\abs{x}+x_0-i\sigma^2k_1)}{2(\sigma^2+\frac{i\hbar t}{2m})}\sum_{n=2s,s=0}^\infty\frac{(-1)^n(n-1)!!\hbar^{5(n+1)}}{(m^2k_0^2\sqrt{2mV_0})^{n+1}2^{n/2}(\sigma^2+\frac{i\hbar t}{2m})^{\frac{n+1}{2}}}\sqrt{\pi}\\ \nonumber 
+\sum_{n=2s-1,s=0}^\infty\frac{\hbar^{5(n+1)}(-1)^nn!!\sqrt{\pi}}{(m^2k_0^2\sqrt{2mV_0})^{n+1}2^{\frac{n+1}{2}}(\sigma^2+\frac{i\hbar t}{2m})^{\frac{n+2}{2}}}]
\end{eqnarray}
\section{Conclusion} We present here the solution of a two state problem where the potential curves are closely placed  or much far apart (In the limit $V_0$ is small and large). The possible corrections over this work if one gives the solution for arbitrary/intermediate values of $V_0$. The possible considerations over this model is absolute value and harmonic potentials which are open problems in this field.

\end{document}